\documentclass[aps,pra,showpacs,amsmath,amssymb,amsfonts,twocolumn,lengthcheck,nofootinbib]{revtex4-1}
\usepackage{graphicx}
\usepackage{subfigure}
\usepackage{verbatim}
\usepackage{dcolumn}
\usepackage{bm}
\usepackage{epsf}
\usepackage{xcolor}
\usepackage{hyperref}
\usepackage{hhline}
\usepackage{txfonts}
\usepackage{float}
\usepackage{enumerate}
\usepackage{bbm}
\usepackage{lipsum}
\usepackage{amsmath}
\usepackage{bbold}
\usepackage{footmisc}
\usepackage{mathrsfs}
\definecolor{mygreen}{rgb}{0.01, 0.31, 0.59}
\definecolor{myblue}{rgb}{0.01, 0.31, 0.59}
\hypersetup{
	colorlinks=true,   
	linkcolor=blue,  
	citecolor=blue, 
	urlcolor =blue    
}

\begin{document}
	
\title{Steady entangled-state generation via cross-Kerr effect in a ferrimagnetic crystal}
\author{Zhi-Bo Yang}
\author{Wei-Jiang Wu}
\author{Jie Li}
\author{Yi-Pu Wang}
\email{yipuwang@zju.edu.cn}
\author{J. Q. You}
\email{jqyou@zju.edu.cn}
\affiliation{Interdisciplinary Center of Quantum Information, State Key Laboratory of Modern Optical Instrumentation, and Zhejiang Province Key Laboratory of Quantum Technology and Device, Department of Physics, Zhejiang University, Hangzhou 310027, China}

\begin{abstract}
For solid-state spin systems, the collective spin motion in a single crystal embodies multiple magnetostatic modes. Recently, it was found that the cross-Kerr interaction between the higher-order magnetostatic mode and the Kittel mode introduces a new operable degree of freedom. In this work, we propose a scheme to entangle two magnon modes via the cross-Kerr nonlinearity when the bias field is inhomogeneous and the system is driven. Quantum entanglement persists at the steady state, as demonstrated by numerical results using experimentally feasible parameters. Furthermore, we also demonstrate that entangled states can survive better in the system where self-Kerr and cross-Kerr nonlinearities coexist. Our work provides insights and guidance for designing experiments to observe entanglement between different degrees of freedom within a single ferrimagnetic crystal. Additionally, it may stimulate potential applications in quantum information processing using spintronic devices.
\end{abstract}
\maketitle

\section{Introduction}
Manipulation of the light-matter interaction has been a long-standing and intriguing topic owing to its important role in quantum information science. In recent years, cavity magnonics has gradually demonstrated its unique advantages when achieving magnon-based hybrid quantum systems~\cite{light-10,HH-13,YT-14,ZX-14,MG-14,bai-15,YC-15,dengke-15,Nakamura-19,Rameshti-21,YuanHY-21}. Among the ferrimagnetic materials and microwave ferrites, the yttrium iron garnet (YIG) has a high spin density ($\sim$$4.22\times10^{27}~{\rm m}^{-3}$) and a low dissipation rate ($\sim$1~MHz). The strong coupling between magnons (quanta of collective spin excitations) in the YIG sphere and cavity photons can be realized, resulting in cavity polaritons~\cite{HH-13,YT-14,ZX-14,MG-14,bai-15}. Moreover, magnons can also interact with visible/near-infrared light waves (via magneto-optical effect~\cite{YRS-1966,RH-16,ZX-16L,AO-16,AO-18}), superconducting qubits (indirectly~\cite{YT-15,YT-16,DLQ-17}), and mechanical deformation modes (directly~\cite{ZX-16,LJ-18,LJ-19,LJ-19njp}) to form various hybrid systems. Experimental and theoretical studies based on cavity magnonics reveal a variety of phenomena, including magnon dark modes~\cite{ZX-15}, magnon Kerr effect~\cite{YP-16,YP-18}, non-Hermitian physics~\cite{MH-17,DZ-17,YP-19,YP-20,Zhao-20,Yang-20}, magnon-induced transparency~\cite{WB-18}, and nonclassical states~\cite{LJ-18,LJ-19,LJ-19njp,ZZZ-19,HYY-20,HYY-20PRB,JMPN-20,ZB-21,ZB-21prr,PRB-2020,WJ-20A,JL-PRXQ}.

Entanglement, as a resource for quantum technologies, plays an essential role in quantum computing~\cite{RR-01,EK-01,Ladd-10,Lidar-18}, quantum metrology~\cite{Lloyd-06,Lloyd-11}, and quantum teleportation~\cite{Pirandola-15}. In addition, it has expanded our understanding of many physical phenomena, such as superradiance~\cite{NL-04}, superconductivity~\cite{VV-04}, and disordered systems~\cite{WD-05}. The mechanism by which continuous variable (CV) entanglement is generated is based on the squeezing-type interactions within the system~\cite{GA-07}. In most systems, this type of interaction is induced by nonlinearities, such as radiation pressure interactions in optomechanical systems~\cite{DV-07,CG-08}, magnetostrictive interactions in cavity magnomechanical systems~\cite{LJ-18,LJ-19njp}, the self-nonlinear Kerr effect in cavity magnonic systems~\cite{ZZZ-19,ZB-21prr}, and other systems that include parametric amplifiers~\cite{GSA-16,LJ-19,JMPN-20,ZB-21}. Also, it exists intrinsically in some particular systems (for instance, the antiferromagnetic system~\cite{AK-19,HYY-20PRB}). The nonlinearity is typically weak (for example, the magnon self-Kerr coefficient has a magnitude of $\sim$0.1 nHz~\cite{RCS-21}), but they can be enhanced by driving the associated spin-wave modes with a drive field.

Cross-Kerr interactions, as a type of nonlinear interaction between fields and waves, have been observed in a variety of systems, including superconducting circuits~\cite{ICH-13,MK-18,AV-20}, atoms~\cite{BH-14,XK-18,JS-19}, and ions~\cite{SD-17}, among others. Recently, the cross-Kerr interaction between the higher-order magnetostatic (HMS) mode and the spin uniform precession mode (referred to as the Kittel mode~\cite{CK-1948}) in cavity magnonics was experimentally observed~\cite{WJ-PRB}. When only one mode is driven, the two spin-wave modes simultaneously undergo nonlinear frequency shift, proving that self- and cross-Kerr effects are simultaneously excited. This nonlinearity enables the formation of entanglement in this system. We anticipate that investigating entanglement properties in such a system is critical for understanding how internal degrees of freedom of the ferrimagnetic crystal are correlated and the effects of different nonlinearities on the entanglement. In light of the importance of producing high-quality entangled photons in quantum computation~\cite{PS-05,OACA-1948,JL-19}, entangled states of magnons may also play a key role in other quantum technologies due to the advantages of controllability, integrability, and reliability in solid-state spin ensembles. In this article, we explore the entanglement between two spin-wave modes inside a single ferrimagnetic crystal based on the cross-Kerr effect.

The article is organized as follows. In Sec.~\ref{s2}, we introduce the fundamental model and derive the effective Hamiltonian. In Sec.~\ref{s3}, the dissipative equations and the covariance matrix of our proposal are given to quantify the bipartite and tripartite entanglements. In Sec.~\ref{s4}, we discuss the cross-Kerr induced entanglement with the optimized effective interaction between the three modes. The condition for optimizing the tripartite entanglement is found and confirmed by the parametric transformation process with four-wave mixing. In Sec.~\ref{s5}, we consider the case where two self-Kerr effects and the cross-Kerr effect coexist. We find that the tripartite entanglement can be enhanced, resulting from the emergence of a new nonlinearity-induced parametric interaction and the enhanced coupling between modes. The numerical results indicate that all entanglements are robust against the temperature changes in the environment. Entanglement detection and application are given in Sec.~\ref{s6}. Finally, we summarize our work in Sec.~\ref{s7}.


\section{The Model Hamiltonian}\label{s2}
\begin{figure}[t]
	\hskip-0.182cm\includegraphics[width=\linewidth]{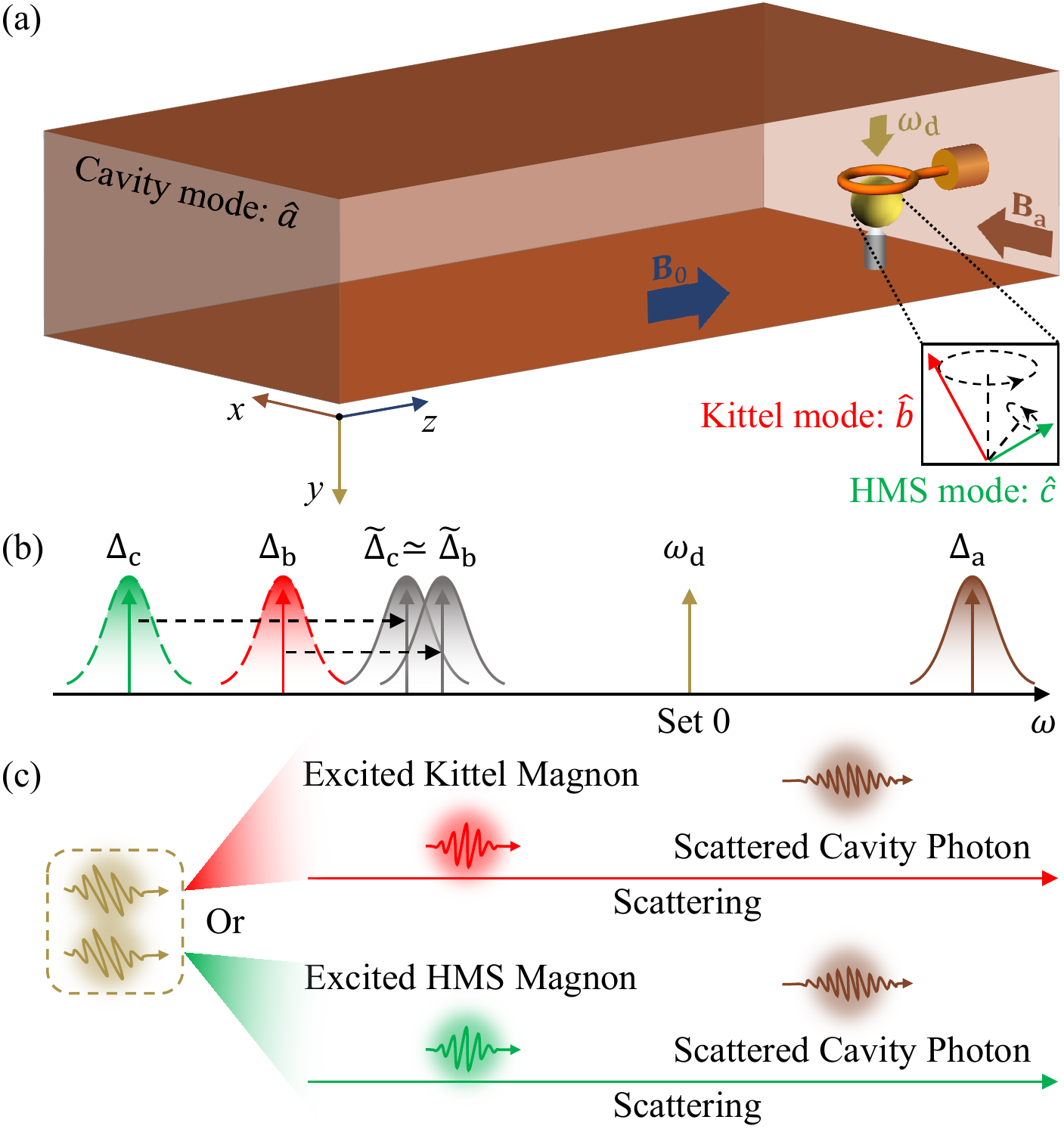}
	\caption{(a) Sketch of the system (the panel is adapted from Ref.~\cite{WJ-PRB}). The YIG sphere is mounted on the cavity wall and magnetized to saturation by a bias magnetic field ${\bf B}_{0}$ along the $z$ direction. The microwave source connected to the loop antenna provides a drive magnetic field, which is aligned along the $y$ direction. The magnetic field of the microwave cavity is in the $x$ direction. Three magnetic fields are mutually perpendicular at the position of the YIG sphere. The two magnetic moments represent the two spin wave modes, respectively [the red (green) arrow corresponds to the Kittel (HMS) mode]. (b) Frequency spectrum of the cavity magnonic system. The existence of self-Kerr and cross-Kerr nonlinearities make the microwave drive induce the effective frequency shift of the two magnon modes. When either the Kittel mode or the HMS mode is driven, the excess energy scatters photons with high frequency. If the cavity photon is matched with the frequency, the system exhibits all bipartite entanglements and genuine tripartite entanglement. (c) Four-wave mixing in cavity magnonics. If the cavity detuning in (b) is tuned to the matching condition ($\Delta_{\rm a}=-\tilde{\Delta}_{\rm b/c}$), a four-wave mixing happens, where two driving photons are adsorbed and the cavity mode and the magnon mode are simultaneously excited.}
	\label{fig1}
\end{figure}


We consider a cavity magnonic system consisting of the cavity mode, Kittel mode, and HMS mode, as shown in Fig.~\ref{fig1}\textcolor{blue}{(a)}. The spin-wave mode interacts with the cavity photon mode via a magnetic dipole-dipole interaction, while the HMS mode couples to the Kittel mode via mode overlap~\cite{AG-19,WJ-PRB,DDS-09}. The Hamiltonian of the whole system reads
\begin{eqnarray}\label{e001}
H=H_{\rm a}+H_{\rm b}+H_{\rm c}+H^{\rm int}_{\rm ab}+H^{\rm int}_{\rm ac}+H^{\rm int}_{\rm bc}+H^{\rm d}_{\rm b}+H^{\rm d}_{\rm c},
\end{eqnarray}
where $H_{\rm a}$, $H_{\rm b}$, and $H_{\rm c}$ describe the free Hamiltonians of the microwave cavity mode, the Kittel mode, and the HMS mode, respectively, $H^{\rm int}_{\rm xy}$ ($xy=ab,ac,bc$) represents the interaction between the corresponding modes, and $H^{\rm d}_{\rm b}$ ($H^{\rm d}_{\rm c}$) represents the coupling between the drive field and the Kittel (HMS) mode. 
It is worth noting that $H_{\rm b}$, $H^{\rm int}_{\rm ab}$, and $H^{\rm d}_{\rm b}$ have similar forms as 
$H_{\rm c}$, $H^{\rm int}_{\rm ac}$, and $H^{\rm d}_{\rm c}$, so derivations of the latter ones will not be repeated hereafter. 

The free Hamiltonian of the cavity mode is 
\begin{eqnarray}\label{e002}
H_{\rm a}=\frac{1}{2}\int\left(\epsilon_{0}\textbf{E}_{\rm a}^2+\frac{\textbf{B}_{\rm a}^2}{\mu_{0}} \right)d\tau,
\end{eqnarray}
where $\textbf{E}_{\rm a}$ ($\textbf{B}_{\rm a}$) is the electric (magnetic) component of the electromagnetic field inside the cavity, and $\epsilon_{0}$ ($\mu_{0}$) is the vacuum permittivity (permeability). By ignoring the constant term, the single-mode electromagnetic field can be quantized as $H_{\rm a}=\hbar\omega_{\rm a}\hat{a}^{\dag}\hat{a}$, with $\hat{a}$ ($\hat{a}^{\dag}$) being the annihilation (creation) operator of the photons at frequency $\omega_{\rm a}$~\cite{DFW-1994}.

The free Hamiltonian of the magnon mode, including the Zeeman energy and the magnetocrystalline anisotropy energy, can be written as~\cite{YP-16}
\begin{eqnarray}\label{e003}
H_{\rm b}=-\int\textbf{M}_{\rm b}\cdot\textbf{B}_{0}d\tau-\frac{\mu_{0}}{2}\int\textbf{M}_{\rm b}\cdot\textbf{H}_{\rm b}^{\rm an}d\tau,
\end{eqnarray}
where $\textbf{B}_{0}=B_{0}\textbf{e}_{\rm z}$ is the applied static magnetic field in the $z$ direction for magnetizing the YIG sphere; $\textbf{e}_{\rm i=x,y,z}$ denote the three orthogonal unit vectors [see Fig.~\ref{fig1}\textcolor{blue}{(a)}]; $\textbf{M}_{\rm b}=\hbar\gamma_{\rm g}\textbf{S}_{\rm b}/V_{\rm yig}\equiv(M^{\rm b}_{\rm x},M^{\rm b}_{\rm y},M^{\rm b}_{\rm z})$ is the magnetization of the Kittel mode in the YIG sphere, with $\gamma_{\rm g}$ being the gyromagnetic ratio ~\cite{light-10}; $V_{\rm yig}$ is the volume of the YIG sphere; and $\textbf{S}_{\rm b}\equiv(S_{\rm x}^{\rm b},S^{\rm b}_{\rm y},S^{\rm b}_{\rm z})$ stands for the collective spin operator of the Kittel mode. When the bias magnetic field is applied along the YIG sphere [100] crystal axis, the anisotropic field is given by~\cite{DDS-09,book1-20}
\begin{eqnarray}\label{e004}
\textbf{H}_{\rm b}^{\rm an}=\frac{2\hbar \gamma_{\rm g}S_{\rm z}^{\rm b}K_{\rm an}^{\rm b}}{\mu_{0}M^2V_{\rm yig}}\textbf{e}_{\rm z},
\end{eqnarray}
where $K_{\rm an}^{\rm b}$ is the dominant first-order anisotropy coefficient and $M$ is the saturation magnetization.

The magnon-photon interaction Hamiltonian is
\begin{eqnarray}\label{e005}
H^{\rm int}_{\rm ab}&=&-\mu_{0}\int\textbf{M}_{\rm b}\cdot\textbf{B}_{\rm a}d\tau,
\end{eqnarray}
where the magnetic field $\textbf{B}_{\rm a}=-[ \hbar\omega_{\rm a}/(\mu_{0}V_{\rm a})] ^\frac{1}{2}(\hat{a}^{\dag}+\hat{a})\textbf{e}_{\rm x}$ of the cavity mode is polarized along the $x$ direction, with $V_{\rm a}$ being the volume of the cavity.

Due to the inhomogeneity of the bias magnetic field $\textbf{B}_{0}$, HMS modes can occur in the YIG sphere. Here we consider one HMS mode near the Kittel mode in frequency. The interaction between the Kittel mode and the HMS mode can be written as
\begin{eqnarray}\label{e006}
H^{\rm int}_{\rm bc}&=&\alpha\int\textbf{M}_{\rm b}\cdot\textbf{M}_{\rm c}d\tau,
\end{eqnarray}
with the coefficient $\alpha$ accounting for the overlap between these two spin-wave modes~\cite{WJ-PRB}. Here, the magnetizations of the Kittel mode and the HMS mode are different ($\textbf{M}_{\rm b}\neq\textbf{M}_{\rm c}$) because more magnetic momentum is contributed to the Kittel mode.   

The interaction Hamiltonian between the drive field and the Kittel mode is
\begin{eqnarray}\label{e007}
H^{\rm d}_{\rm b}&=&-\mu_{0}\int\textbf{M}_{\rm b}\cdot\textbf{H}_{\rm d}d\tau,
\end{eqnarray}
where $\textbf{H}_{\rm d}=iB_{\rm d}\cos(\omega_{\rm d}t)\textbf{e}_{\rm y}$ represents the drive field along the $y$ direction, with drive frequency (amplitude) $\omega_{\rm d}$ ($B_{\rm d}$). For the low-lying magnon excitations with $\langle \hat{o}^{\dag}\hat{o} \rangle\ll2S_{\rm o}$, where $o=b,c$, the Holstein-Primakoff transformations of the two modes are given by~\cite{HP-1940}
\begin{eqnarray}\label{e008}
S_{\rm o}^{\rm z}&=&S_{\rm o}-\hat{o}^{\dag}\hat{o},\nonumber\\
S_{\rm o}^{+}&=&\hat{o}(2S_{\rm o}-\hat{o}^{\dag}\hat{o})^{\frac{1}{2}}\simeq(2S_{\rm o})^{\frac{1}{2}}\hat{o},\nonumber\\
S_{\rm o}^{-}&=&\hat{o}^{\dag}(2S_{\rm o}-\hat{o}^{\dag}\hat{o})^{\frac{1}{2}}\simeq(2S_{\rm o})^{\frac{1}{2}}\hat{o}^{\dag},
\end{eqnarray}
with $S_{\rm o}^{\pm}\equiv S_{\rm o}^{\rm x}\pm iS_{\rm o}^{\rm y}$.

Under the rotating-wave approximation~\cite{DFW-1994}, the total effective Hamiltonian of the cavity magnonic system can be rewritten as
\begin{eqnarray}\label{e009}
H/\hbar&=&\omega_{\rm a}\hat{a}^\dag \hat{a}+\omega_{\rm b}\hat{b}^\dag \hat{b}+\omega_{\rm c}\hat{c}^\dag \hat{c}+K_{\rm b}\hat{b}^\dag \hat{b}\hat{b}^\dag \hat{b}+K_{\rm c}\hat{c}^\dag \hat{c}\hat{c}^\dag \hat{c}\nonumber\\
&&+g_{\rm ab}(\hat{a}^\dag\hat{b}+\hat{a}\hat{b}^\dag)+g_{\rm ac}(\hat{a}^\dag\hat{c}+\hat{a}\hat{c}^\dag)+g_{\rm bc}(\hat{b}^\dag\hat{c}+\hat{b}\hat{c}^\dag)\nonumber\\
&&+G\hat{b}^\dag \hat{b}\hat{c}^\dag \hat{c}+\Omega_{\rm b}(\hat{b}^\dag e^{-i\omega_{\rm d}t}-\hat{b}e^{i\omega_{\rm d}t})\nonumber\\
&&+\Omega_{\rm c}(\hat{c}^\dag e^{-i\omega_{\rm d}t}-\hat{c}e^{i\omega_{\rm d}t}),
\end{eqnarray}
where
\begin{eqnarray}\label{e010}
\omega_{\rm b(c)}=\gamma_{\rm g}B_{0}-\frac{2\hbar\gamma_{\rm g}^2 K_{\rm an}^{\rm b(c)}S_{\rm b(c)}}{M^{2}V_{\rm yig}}-\frac{\alpha\hbar\gamma_{\rm g}^2S_{\rm c(b)}}{V_{\rm yig}}
\end{eqnarray}
is the angular frequency of the Kittel (HMS) mode,
\begin{eqnarray}\label{e011}
K_{\rm b(c)}=-\frac{\hbar\gamma_{\rm g}^2K_{\rm an}^{\rm b(c)}}{M^2V_{\rm yig}}~{\rm with}~K_{\rm an}^{\rm b(c)}<0
\end{eqnarray}
is the self-Kerr nonlinear coefficients of the Kittel (HMS) mode, and
\begin{eqnarray}\label{e012}
g_{\rm ab(ac)}&=&\sqrt{\frac{S_{\rm b(c)}\gamma_{\rm g}^2\mu_{0}\hbar\omega_{\rm a}}{2V_{\rm a}}}
\end{eqnarray}
denotes the coupling strength between the cavity mode and the Kittel (HMS) mode. Moreover, $g_{\rm bc}=\alpha\hbar\gamma_{\rm g}^2(S_{\rm b}S_{\rm c})^{\frac{1}{2}}/V_{\rm yig}$ represents the coupling between the two magnon modes, $G=\alpha\hbar\gamma_{\rm g}^2/V_{\rm yig}$ is the cross-Kerr coefficient, and $\Omega_{\rm b(c)}=\frac{1}{4}\mu_{0}\gamma_{\rm g}B_{\rm d}(2S_{\rm b(c)})^{\frac{1}{2}}$ are the Rabi frequencies of the two spin-wave modes. 

For the Kittel mode in a micrometer-scale YIG sphere, its spin moment contributes dominantly to the dipole than that of the MHS mode~\cite{WJ-PRB}. Thus, we can reasonably ignore the coupling between the cavity mode and the HMS mode. Similarly, the beam-splitter-type interaction between the Kittel mode and the HMS mode is also small and can hence be neglected in the analysis. Then, in the rotating frame with respect to the drive frequency $\omega_{\rm d}$, the system Hamiltonian can be reduced to
\begin{eqnarray}\label{e013}
H_{\rm eff}/\hbar&=&\Delta_{\rm a}\hat{a}^\dag \hat{a}+\Delta_{\rm b}\hat{b}^\dag \hat{b}+\Delta_{\rm c}\hat{c}^\dag \hat{c}+K_{\rm b}\hat{b}^\dag \hat{b}\hat{b}^\dag \hat{b}+K_{\rm c}\hat{c}^\dag \hat{c}\hat{c}^\dag \hat{c}\nonumber\\
&&+g_{\rm ab}(\hat{a}^\dag\hat{b}+\hat{a}\hat{b}^\dag)+G\hat{b}^\dag \hat{b}\hat{c}^\dag \hat{c}+\Omega_{\rm b}(\hat{b}^\dag-\hat{b})\nonumber\\
&&+\Omega_{\rm c}(\hat{c}^\dag-\hat{c}),
\end{eqnarray}
where $\Delta_{\rm a (b,c)}=\omega_{\rm a (b,c)}-\omega_{\rm d}$.

\section{Dissipative Equations and Covariance Matrix}\label{s3}
Due to the coupling between the cavity magnonic system and the environment, the system will inevitably be influenced by the cavity decay and magnetic damping. Taking these dissipative elements into account, the dissipative dynamics of the system is described by a set of quantum Langevin equations (QLEs)
\begin{eqnarray}\label{e014}
\frac{d\hat{a}}{dt}&=&-(i\Delta_{\rm a}+\gamma_{\rm a})\hat{a}-ig_{\rm ab}\hat{b}+(2\gamma_{\rm a})^\frac{1}{2}\hat{a}^{\rm in},\\
\frac{d\hat{b}}{dt}&=&-(i\Delta_{\rm b}+\gamma_{\rm b})\hat{b}-ig_{\rm ab}\hat{a}-i\Omega_{\rm b}-iG\hat{b}\hat{c}^\dag \hat{c}+(2\gamma_{\rm b})^\frac{1}{2}\hat{b}^{\rm in}\nonumber\\
&&-2iK_{\rm b}\hat{b}^\dag \hat{b}\hat{b},\nonumber\\
\frac{d\hat{c}}{dt}&=&-(i\Delta_{\rm c}+\gamma_{\rm c})\hat{c}-2iK_{\rm c}\hat{c}^\dag \hat{c}\hat{c}-iG\hat{c}\hat{b}^\dag \hat{b}-i\Omega_{\rm c}+(2\gamma_{\rm c})^\frac{1}{2}\hat{c}^{\rm in},\nonumber
\end{eqnarray}
where $\gamma_{\rm a}$, $\gamma_{\rm b}$, and $\gamma_{\rm c}$ ($\hat{a}^{\rm in}$, $\hat{b}^{\rm in}$, and $\hat{c}^{\rm in}$) represent the damping rates (the zero-mean input noise operators) of the cavity mode, the Kittel mode, and the HMS mode, respectively. Under the Markovian reservoir assumption, the input noise operators are characterized by the following correlation functions~\cite{CWG-2004}:
\begin{eqnarray}\label{e015}
\langle \hat{o}^{\rm in\dag}(t)\hat{o}^{\rm in}(t^{\prime})\rangle&=&n_{\rm o}\delta(t-t^{\prime}),\nonumber\\
\langle \hat{o}^{\rm in}(t)\hat{o}^{\rm in\dag}(t^{\prime})\rangle&=&(n_{\rm o}+1)\delta(t-t^{\prime}),
\end{eqnarray}
where $n_{\rm o}=\{\exp[\hbar\omega_{\rm o}/(k_{\rm B}T_{\rm e})]-1\}^{-1}$ being equilibrium mean thermal photon ($o=a$) and magnon ($o=b,c$) numbers. $T_{\rm e}$ is the environmental temperature, and $k_{\rm B}$ is the Boltzmann constant. Because the YIG crystal is strongly driven by a microwave field, the couplings between different modes are unhindered, resulting in these modes all having large amplitudes (i.e., $\arrowvert\langle o\rangle\arrowvert\gg1$), so the standard linearization treatment can be applied to the nonlinear QLEs~[Eq.~(\ref{e014})]. In this case, one can safely introduce the expansion $\hat{o}=\langle o\rangle+o$ in the vicinity of steady-state averages by neglecting higher-order fluctuations of the operators. Then, we obtain a set of differential equations for mean values
\begin{eqnarray}\label{e016}
\frac{d\langle a\rangle}{dt}&=&-(i\Delta_{\rm a}+\gamma_{\rm a})\langle a\rangle-ig_{\rm ab}\langle b\rangle,\\
\frac{d\langle b\rangle}{dt}&=&-(i\Delta_{\rm b}+\gamma_{\rm b})\langle b\rangle-2iK_{\rm b}\arrowvert\langle b\rangle \arrowvert^{2}\langle b\rangle-iG\arrowvert\langle c\rangle \arrowvert^{2}\langle b\rangle\nonumber\\
&&-ig_{\rm ab}\langle a\rangle-i\Omega_{\rm b},\nonumber\\
\frac{d\langle c\rangle}{dt}&=&-(i\Delta_{\rm c}+\gamma_{\rm c})\langle c\rangle-2iK_{\rm c}\arrowvert\langle c\rangle \arrowvert^{2}\langle c\rangle-iG\arrowvert\langle b\rangle \arrowvert^{2}\langle c\rangle-i\Omega_{\rm c}.\nonumber
\end{eqnarray}
The steady state solution of the system satisfies the following equations
\begin{eqnarray}\label{e017}
0&=&-(i\Delta_{\rm b}+\gamma_{\rm b})\langle b\rangle-2iK_{\rm b}\arrowvert\langle b\rangle \arrowvert^{2}\langle b\rangle-iG\arrowvert\langle c\rangle \arrowvert^{2}\langle b\rangle-i\Omega_{\rm b}\nonumber\\
&&-g_{\rm ab}^2\langle b\rangle/(i\Delta_{\rm a}+\gamma_{\rm a}),\\
0&=&-(i\Delta_{\rm c}+\gamma_{\rm c})\langle c\rangle-2iK_{\rm c}\arrowvert\langle c\rangle \arrowvert^{2}\langle c\rangle-iG\arrowvert\langle b\rangle \arrowvert^{2}\langle c\rangle-i\Omega_{\rm c}.\nonumber
\end{eqnarray}
Supposing $\arrowvert\Delta_{\rm a}\arrowvert\gg\gamma_{\rm a}$, $\arrowvert\Delta_{\rm b}\arrowvert\gg\gamma_{\rm b}$, and $\arrowvert\Delta_{\rm c}\arrowvert\gg\gamma_{\rm c}$, we can
approximately get that $\langle a\rangle$, $\langle b\rangle$, and $\langle c\rangle$ are all pure real numbers. It should be noted that the approximation is only used to demonstrate that $\langle a\rangle$, $\langle b\rangle$, and $\langle c\rangle$ are approximately real numbers, which simplify the following calculations. However, the damping terms are included in all subsequent calculations and numerical simulations, which are necessary for the system to reach the steady state.

The linearized QLEs for the quantum fluctuations can be written as
\begin{eqnarray}\label{e019}
\frac{da}{dt}&=&-(i\Delta_{\rm a}+\gamma_{\rm a})a-ig_{\rm ab}b+(2\gamma_{\rm a})^\frac{1}{2}a^{\rm in},\\
\frac{db}{dt}&=&-(i\tilde{\Delta}_{\rm b}+\gamma_{\rm b})b-i\tilde{K}_{\rm b}b^{\dag}-ig_{\rm ab}a-i\tilde{G}(c^{\dag}+c)\nonumber\\
&&+(2\gamma_{\rm b})^\frac{1}{2}b^{\rm in},\nonumber\\
\frac{dc}{dt}&=&-(i\tilde{\Delta}_{\rm c}+\gamma_{\rm c})c-i\tilde{K}_{\rm c}c^{\dag}-i\tilde{G}(b^{\dag}+b)+(2\gamma_{\rm c})^\frac{1}{2}c^{\rm in},\nonumber
\end{eqnarray}
where $\tilde{\Delta}_{\rm b}=\Delta_{\rm b}+4K_{\rm b}\arrowvert\langle b\rangle\arrowvert^2+G\arrowvert\langle c\rangle\arrowvert^2$ and $\tilde{\Delta}_{\rm c}=\Delta_{\rm c}+4K_{\rm c}\arrowvert\langle c\rangle\arrowvert^2+G\arrowvert\langle b\rangle\arrowvert^2$ are the effective magnon mode-drive field detunings including the frequency shifts caused by the self-Kerr and cross-Kerr effects. $\tilde{K}_{\rm b}=2K_{\rm b}\langle b\rangle^2$ and $\tilde{K}_{\rm c}=2K_{\rm c}\langle c\rangle^2$ are effective self-Kerr coefficients, and $\tilde{G}=G\langle b\rangle\langle c\rangle$ is the effective magnon-magnon coupling rate. Since $\langle b\rangle$ and $\langle c\rangle$ are approximately pure real numbers, then $\tilde{K}_{\rm b}\simeq2K_{\rm b}\arrowvert\langle b\rangle\arrowvert^2$ and $\tilde{K}_{\rm c}\simeq2K_{\rm c}\arrowvert\langle c\rangle\arrowvert^2$. In this case, we have $\tilde{\Delta}_{\rm b}\simeq\Delta_{\rm b}+2\tilde{K}_{\rm b}+G\arrowvert\langle c\rangle\arrowvert^2$ and $\tilde{\Delta}_{\rm c}\simeq\Delta_{\rm c}+2\tilde{K}_{\rm c}+G\arrowvert\langle b\rangle\arrowvert^2$.

In order to study the quantum correlation induced by the cross-Kerr effect, we first need to determine the cross-Kerr coefficient $G$. This can be done by using the parameters reported in the recent experiment~\cite{WJ-PRB}. Due to the large frequency detuning between the Kittel mode and the HMS mode, the HMS mode is not excited when only the Kittel mode is driven (i.e., $\Omega_{\rm b}\not=0$, $\Omega_{\rm c}=0$). In this case, the frequency shift ($2K_{\rm b}\arrowvert\langle b\rangle \arrowvert^{2}/2\pi\simeq-60$ MHz) of the Kittel mode is caused by the self-Kerr nonlinearity, while the frequency shift ($G\arrowvert\langle b\rangle \arrowvert^{2}/2\pi\simeq-150$ MHz) of the HMS mode is due to the cross-Kerr effect. Therefore, $K_{\rm b}/2\pi\simeq-0.1$ nHz (the bias magnetic field is applied along the YIG sphere [110] crystal axis) corresponds to $G/2\pi\simeq-0.5$ nHz can be found in a 1-mm-diameter YIG sphere. Similarly, the Kittel mode is not excited when only the HMS mode is driven (i.e., $\Omega_{\rm c}\not=0$, $\Omega_{\rm b}=0$). The frequency shift ($2K_{\rm c}\arrowvert\langle c\rangle \arrowvert^{2}/2\pi\simeq-24$ MHz) of the HMS mode is caused by the self-Kerr nonlinearity, while the frequency shift ($G\arrowvert\langle c\rangle \arrowvert^{2}/2\pi\simeq-10$ MHz) of the Kittel mode is due to the cross-Kerr effect. So we can get $K_{\rm c}/2\pi\simeq-0.6$ nHz.

In this work, we consider the following situation: the bias magnetic field is applied along the YIG sphere [100] crystal axis (i.e., $K_{\rm b(c)},G>0$) and two drive fields are applied at the same time ($\Omega_{\rm b(c)}\not=0$), where $2K_{\rm b}\arrowvert\langle b\rangle \arrowvert^{2}/2\pi\simeq15$ MHz and $G\arrowvert\langle b\rangle \arrowvert^{2}/2\pi\simeq37.5$ MHz corresponds to $\arrowvert\langle b\rangle \arrowvert^{2}\simeq7.5\times10^{16}$; $2K_{\rm c}\arrowvert\langle c\rangle \arrowvert^{2}/2\pi\simeq24$ MHz and $G\arrowvert\langle c\rangle \arrowvert^{2}/2\pi\simeq10$ MHz corresponds to $\arrowvert\langle c\rangle \arrowvert^{2}\simeq2\times10^{16}$. In this case, we have $\tilde{G}/2\pi\simeq19.4$ MHz, which is utilized in the following analysis. More parameter details are listed in Tab.~\ref{tab1}.

\begin{figure*}[t]
	\hskip-0.19cm\includegraphics[width=\linewidth]{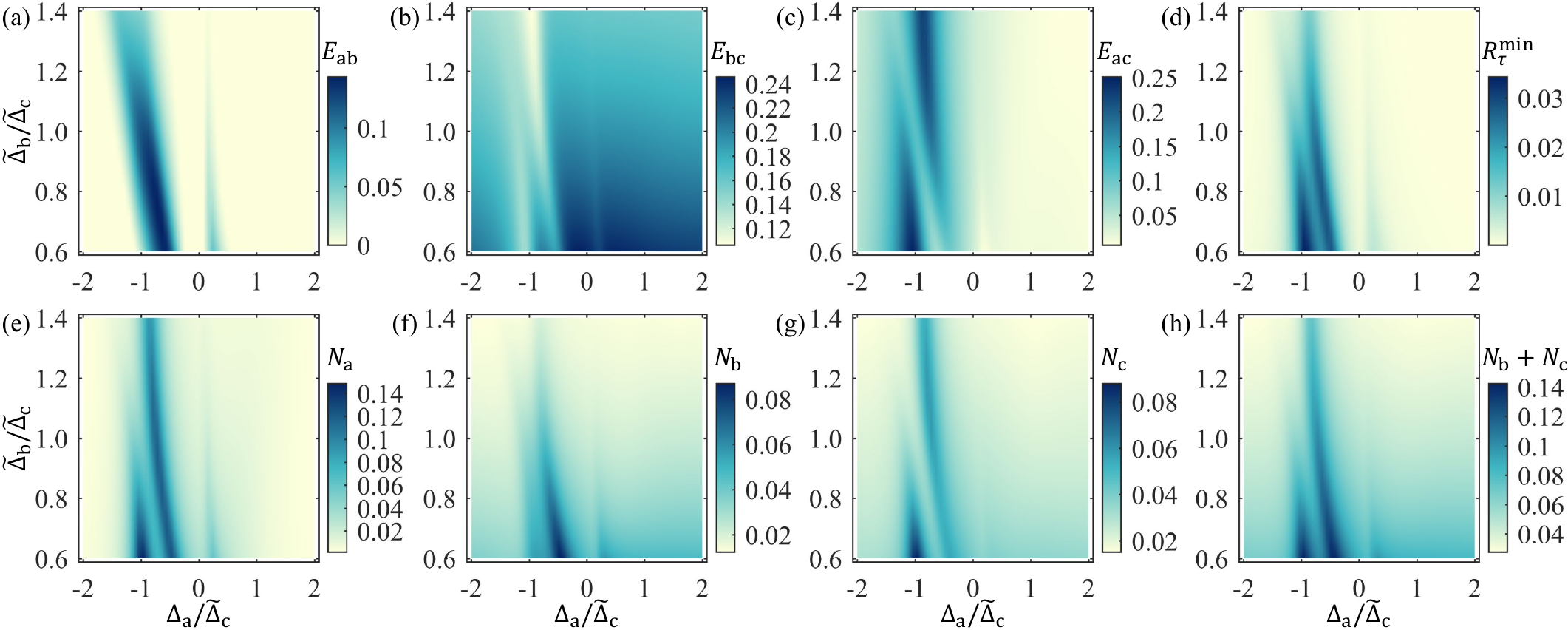}
	\caption{Bipartite entanglements (a) $E_{\rm ab}$, (b) $E_{\rm bc}$, (c) $E_{\rm ac}$ and (d) tripartite entanglement $R_{\tau}^{\rm min}$ versus detunings $\tilde{\Delta}_{\rm b}$ and $\Delta_{\rm a}$. Mean excitation numbers for (e) the cavity mode $N_{\rm a}$, (f) the Kittel mode $N_{\rm b}$, (g) the HMS mode $N_{\rm c}$, and (h) the sum of $N_{\rm b}$ and $N_{\rm c}$ versus $\tilde{\Delta}_{\rm b}$ and $\Delta_{\rm a}$. See Tab.~\ref{tab1} for the detailed parameters.}
	\label{fig2}
\end{figure*}
To quantify the entanglement of the system, we introduce the quadrature fluctuation (noise) operators: $X_{\rm o}=(o+o^{\dag})/\sqrt{2}$, $Y_{\rm o}=i(o^{\dag}-o)/\sqrt{2}$, $X_{\rm o}^{\rm in}=(o^{\rm in}+o^{\rm in\dag})/\sqrt{2}$, and $Y_{\rm o}^{\rm in}=i(o^{\rm in\dag}-o^{\rm in})/\sqrt{2}$ with $o=a,b,c$. The linearized QLEs~[Eq.~(\ref{e019})] for the quadrature fluctuations can be writen as
\begin{eqnarray}\label{e020}
\frac{dX_{\rm a}}{dt}&=&\Delta_{\rm a}Y_{\rm a}-\gamma_{\rm a}X_{\rm a}+g_{\rm ab}Y_{\rm b}+(2\gamma_{\rm a})^\frac{1}{2}X_{\rm a}^{\rm in},\nonumber\\
\frac{dY_{\rm a}}{dt}&=&-\Delta_{\rm a}X_{\rm a}-\gamma_{\rm a}Y_{\rm a}-g_{\rm ab}X_{\rm b}+(2\gamma_{\rm a})^\frac{1}{2}Y_{\rm a}^{\rm in},\nonumber\\
\frac{dX_{\rm b}}{dt}&=&(\tilde{\Delta}_{\rm b}-\tilde{K}_{\rm b})Y_{\rm b}-\gamma_{\rm b}X_{\rm b}+g_{\rm ab}Y_{\rm a}+(2\gamma_{\rm b})^\frac{1}{2}X_{\rm b}^{\rm in},\nonumber\\
\frac{dY_{\rm b}}{dt}&=&-(\tilde{\Delta}_{\rm b}+\tilde{K}_{\rm b})X_{\rm b}-\gamma_{\rm b}Y_{\rm b}-g_{\rm ab}X_{\rm a}-2\tilde{G}X_{\rm c}+(2\gamma_{\rm b})^\frac{1}{2}Y_{\rm b}^{\rm in},\nonumber\\
\frac{dX_{\rm c}}{dt}&=&(\tilde{\Delta}_{\rm c}-\tilde{K}_{\rm c})Y_{\rm c}-\gamma_{\rm c}X_{\rm c}+(2\gamma_{\rm c})^\frac{1}{2}X_{\rm c}^{\rm in},\nonumber\\
\frac{dY_{\rm c}}{dt}&=&-(\tilde{\Delta}_{\rm c}+\tilde{K}_{\rm c})X_{\rm c}-\gamma_{\rm c}Y_{\rm c}-2\tilde{G}X_{\rm b}+(2\gamma_{\rm c})^\frac{1}{2}Y_{\rm c}^{\rm in}.
\end{eqnarray}
Then Eq.~(\ref{e020}) can be expressed in a more concise form:
\begin{eqnarray}\label{e021}
\frac{du}{dt}=Au(t)+v(t),
\end{eqnarray}
where the drift matrix $A$ reads
\begin{eqnarray}\label{e022}
A =
\begin{pmatrix}
-\gamma_{\rm a}&\Delta_{\rm a}&  0 &g_{\rm ab}&  0  &0   \\
-\Delta_{\rm a}&-\gamma_{\rm a}&-g_{\rm ab}& 0  &0&  0   \\
0 &g_{\rm ab}&-\gamma_{\rm b}&\tilde{\Delta}_{\rm b}-\tilde{K}_{\rm b}&0&0 \\
-g_{\rm ab}& 0 & -\tilde{\Delta}_{\rm b}-\tilde{K}_{\rm b}&-\gamma_{\rm b} &-2\tilde{G}&  0 \\
0 &0  &  0  & 0 &  -\gamma_{\rm c}  & \tilde{\Delta}_{\rm c}-\tilde{K}_{\rm c}  \\
0 &  0  &  -2\tilde{G}  &  0  & -\tilde{\Delta}_{\rm c}-\tilde{K}_{\rm c} & -\gamma_{\rm c}  \\
\end{pmatrix},\nonumber\\
\end{eqnarray}
and $u=[X_{\rm a}$, $Y_{\rm a}$, $X_{\rm b}$, $Y_{\rm b}$, $X_{\rm c}$, $Y_{\rm c}]^T$, $v=[(2\gamma_{\rm a})^\frac{1}{2}X_{\rm a}^{\rm in}$, $(2\gamma_{\rm a})^\frac{1}{2}Y_{\rm a}^{\rm in}$, $(2\gamma_{\rm b})^\frac{1}{2}X_{\rm b}^{\rm in}$, $(2\gamma_{\rm b})^\frac{1}{2}Y_{\rm b}^{\rm in}$, $(2\gamma_{\rm c})^\frac{1}{2}X_{\rm c}^{\rm in}$, $(2\gamma_{\rm c})^\frac{1}{2}Y_{\rm c}^{\rm in}]^T$ are, respectively, the vectors for quantum fluctuations and noises.
\begin{table}[b]
	\caption{\label{tab1}%
		Description of experimentally feasible parameters. For related parameter detection, please refer to Refs.~\cite{YP-16,YP-18,YP-19,WJ-PRB}.}
	\begin{ruledtabular}
		\begin{tabular}{cc}
			\multicolumn{1}{c}{\textrm{Parameters:}}&
			\multicolumn{1}{c}{\textrm{Value (all figures):}}\\
			\hline
			\mbox{$\omega_{\rm a}/(2\pi\times{\rm GHz})$}&Figs.~\ref{fig2}-\ref{fig4}: 10.07\\
			\mbox{$\omega_{\rm b}/(2\pi\times{\rm GHz})$}&Figs.~\ref{fig2}-\ref{fig4}: 9.86\\
			\mbox{$\omega_{\rm c}/(2\pi\times{\rm GHz})$}&Figs.~\ref{fig2}-\ref{fig4}: 9.7845\\
			\mbox{$\omega_{\rm d}/(2\pi\times{\rm GHz})$}&Figs.~\ref{fig2}-\ref{fig4}: 9.97\\
			\mbox{$\Delta_{\rm a}/(2\pi\times{\rm MHz})$}&Fig.~\ref{fig4}: 100\\
			\mbox{$\Delta_{\rm b}/(2\pi\times{\rm MHz})$}&Figs.~\ref{fig2}-\ref{fig4}: $-$110\\
			\mbox{$\Delta_{\rm c}/(2\pi\times{\rm MHz})$}&Figs.~\ref{fig2}-\ref{fig4}: $-$185.5\\
			\mbox{$g_{\rm ab}/(2\pi\times{\rm MHz})$}&Fig.~\ref{fig2}: 35; Figs.~\ref{fig3}-\ref{fig4}: 30\\
			\mbox{$\gamma_{\rm a}/(2\pi\times{\rm MHz})$}&Fig.~\ref{fig2}: 5.5; Figs.~\ref{fig3}-\ref{fig4}: 18.6\\
			\mbox{$\gamma_{\rm b}/(2\pi\times{\rm MHz})$}&Fig.~\ref{fig2}: 12; Figs.~\ref{fig3}-\ref{fig4}: 6.7\\
			\mbox{$\gamma_{\rm c}/(2\pi\times{\rm MHz})$}&Fig.~\ref{fig2}: 12; Figs.~\ref{fig3}-\ref{fig4}: 6.7\\
			\mbox{$K_{\rm b}/(2\pi\times{\rm nHz})$}&$0.1$ ([100] crystal axis)\\
			\mbox{$K_{\rm c}/(2\pi\times{\rm nHz})$}&$0.6$ ([100] crystal axis)\\
			\mbox{$G/(2\pi\times{\rm nHz})$}&$0.5$ ([100] crystal axis)\\
			\mbox{$\tilde{K}_{\rm b}/(2\pi\times{\rm MHz})$}&Fig.~\ref{fig2}: 0; Fig.~\ref{fig3}: 0, 7.5, 15; Fig.~\ref{fig4}: 15\\
			\mbox{$\tilde{K}_{\rm c}/(2\pi\times{\rm MHz})$}&Fig.~\ref{fig2}: 0; Fig.~\ref{fig3}: 0, 12, 24; Fig.~\ref{fig4}: 24\\
			\mbox{$\tilde{\Delta}_{\rm b}/(2\pi\times{\rm MHz})$}&Figs.~\ref{fig3}-\ref{fig4}: $-$70\\
			\mbox{$\tilde{\Delta}_{\rm c}/(2\pi\times{\rm MHz})$}&Figs.~\ref{fig2}-\ref{fig4}: $-$100\\
			\mbox{$\tilde{G}/(2\pi\times{\rm MHz})$}&Figs.~\ref{fig2}-\ref{fig4}: 19.4\\
			\mbox{$T_{\rm e}/{\rm Kelvin}$}&Figs.~\ref{fig2}-\ref{fig3}: 0; Fig.~\ref{fig4}: 0.15$\sim$0.2 (MST)\\
		\end{tabular}
	\end{ruledtabular}
\end{table}
Since the dynamics of the system is governed by a set of linearized QLEs, the Gaussian nature of the input states will be preserved during the time evolution. That is, the steady state of the quantum fluctuations of the system is a CV three-mode Gaussian state. The state can be fully characterized by a stationary covariance matrix (CM) $V$ whose matrix element is defined by
\begin{eqnarray}\label{e023}
V_{\rm ij}=\frac{1}{2}\langle u_{\rm i}(t)u_{\rm j}(t^{\prime})+u_{\rm j}(t^{\prime})u_{\rm i}(t)\rangle
\end{eqnarray}
with $i,j=1,2,\dots,6$.
The matrix $V$ is obtained by solving the Lyapunov equation~\cite{PCP-1993,DV-07}
\begin{eqnarray}\label{e024}
\frac{dV}{dt}=A(t)V(t)+V(t)A^{T}(t)+D,
\end{eqnarray}
where $D={\rm diag}[\gamma_{\rm a}(2n_{\rm a}+1)$, $\gamma_{\rm a}(2n_{\rm a}+1)$, $\gamma_{\rm b}(2n_{\rm b}+1)$, $\gamma_{\rm b}(2n_{\rm b}+1)$, $\gamma_{\rm c}(2n_{\rm c}+1)$, $\gamma_{\rm c}(2n_{\rm c}+1)]$ is a diffusion matrix, and whose matrix element is related to the noise correlations and defined by
\begin{eqnarray}\label{e025}
D_{\rm ij}=\frac{\langle v_{\rm i}(t)v_{\rm j}(t^{\prime})+v_{\rm j}(t^{\prime})v_{\rm i}(t)\rangle}{2\delta(t-t^{\prime})}.
\end{eqnarray}

To study the bipartite CV entanglements, we introduce the logarithmic negativity $E_{\rm N}$ (which includes $E_{\rm ab}$: cavity-Kittel entanglement, $E_{\rm ac}$: cavity-HMS entanglement, and $E_{\rm bc}$: Kittel-HMS entanglement). And for the tripartite entanglement, we use the minimum residual contangle $R^{\rm min}_{\tau}$, of which the definitions can be found in the Appendix~\ref{appA}, which includes Refs.~\cite{GV-02,MBP-05,GA-06,GA-07}.

\section{Entangled State Generation via cross-Kerr effect}\label{s4}
In order to show the behavior of the entanglements induced by the cross-Kerr effect, Figs.~\ref{fig2}\textcolor{blue}{(a)}-\ref{fig2}\textcolor{blue}{(d)} describe bipartite entanglements $E_{\rm ab}$~[Fig.~\ref{fig2}\textcolor{blue}{(a)}], $E_{\rm ac}$~[Fig.~\ref{fig2}\textcolor{blue}{(b)}], $E_{\rm bc}$~[Fig.~\ref{fig2}\textcolor{blue}{(c)}] and tripartite entanglement~[Fig.~\ref{fig2}\textcolor{blue}{(d)}] $R_{\tau}^{\rm min}$ as a function of two detunings $\tilde{\Delta}_{\rm b}$ and $\Delta_{\rm a}$, where the two self-Kerr effects are not considered. In fact, the self-Kerr effects of the two modes can be characterized by their magnetocrystalline anisotropy energies, which are defined as~\cite{RCS-21}
\begin{eqnarray}\label{e030}
	\mathcal{E}_{\rm b(c)}/\hbar=-\frac{2 \gamma_{\rm g}S_{\rm z}^{\rm b(c)}M^{\rm b(c)}_{\rm z}K_{\rm an}^{\rm b(c)}}{M^2}.
\end{eqnarray}
The energies may be eliminated by adjusting the angle between the bias magnetic field and the crystal axis~\cite{book1-20}. According to the Routh-Hurwitz criterion, the system is stable and reaches its steady state when all the eigenvalues of the matrix $A$ have negative real parts. Therefore, we start our analysis by determining the eigenvalues of the matrix $A$ (i.e., $\arrowvert A-\lambda\textbf{I} \arrowvert=0$) and make sure that the stability conditions are all satisfied in numerical simulations. The parameters used in the article are shown in Tab.~\ref{tab1}, which are adopted from the experimental studies: Refs.~\cite{YP-16,YP-18,YP-19,WJ-PRB}.

As illustrated in Fig~\ref{fig2}\textcolor{blue}{(b)}, the entanglement $E_{\rm bc}$ emerges when $\tilde{\Delta}_{\rm b}\simeq\tilde{\Delta}_{\rm c}$. However, when the spin wave subsystems are coupled to the cavity field near-resonantly, that is, when $\Delta_{\rm a}\simeq-\tilde{\Delta}_{\rm c}$, $E_{\rm bc}$ is partially transferred to cavity mode-Kittel mode and cavity mode-HMS mode subsystems. As a result, entanglements $E_{\rm ab}$ and $E_{\rm ac}$ arise, as shown in Figs.~\ref{fig2}\textcolor{blue}{(a)} and~\ref{fig2}\textcolor{blue}{(c)}. At the same time, the tripartite entanglement of the system can be generated [see Fig.~\ref{fig2}\textcolor{blue}{(d)}]. The physics behind the scenes are as follows: the two magnon modes (the Kittel mode and the HMS mode) are initially detuned and can be driven strongly by the microwave source. Due to the existence of the self-Kerr and cross-Kerr effects, two magnon modes can be driven close to resonance. We begin by demonstrating the situation in the absence of two self-Kerr effects, which is necessary for elucidating the condition for optimizing magnon-magnon entanglement solely through cross-Kerr nonlinearity. Then, we proceed the analysis of the quantum fluctuations via the linearized Hamiltonian
\begin{eqnarray}\label{e031}
	H_{\rm flu}/\hbar&=&\Delta_{\rm a}a^\dag a+\tilde{\Delta}_{\rm b}b^\dag b+\tilde{\Delta}_{\rm c}c^\dag c+\tilde{G}(b^\dag+b)(c^\dag+c)\nonumber\\
	&&+g_{\rm ab}(a^\dag b+ab^\dag),
\end{eqnarray}
where $\tilde{G}(b^\dag c^\dag+bc)$ implies the two-mode-squeezing-type interaction between the Kittel mode and the HMS mode induced by the cross-Kerr effect, which can be significantly enhanced by driving the magnon modes. The Hamiltonian~(\ref{e031}) describes the magnon-magnon entanglement when $\tilde{G}\not=0$. If the cavity field is further participated in the entanglement production and scattering, the four-wave mixing gives rises to the magnon mode-cavity mode entanglement [see Figs.~\ref{fig1}\textcolor{blue}{(b)} and \ref{fig1}\textcolor{blue}{(c)}]. The spontaneous parametric process leads to the transfer of entanglement at suitable detuning frequencies (i.e., matching condition: $\Delta_{\rm a}\simeq-\tilde{\Delta}_{\rm b/c}$). In this case, the indirectly coupled cavity photons and HMS mode magnons get entangled and the entanglement is even larger than those in directly coupled subsystems [see Figs.~\ref{fig2}\textcolor{blue}{(b)} and \ref{fig2}\textcolor{blue}{(c)} for $\Delta_{\rm a}=-\tilde{\Delta}_{\rm b/c}$]. Similar mechanism with three-wave mixing has also been found in optomechanical systems~\cite{CG-08} and cavity magnetomechanical systems~\cite{LJ-18}.

To demonstrate the conversion process, we introduce the final mean photon and magnon numbers, which can be calculated by the relation
\begin{eqnarray}\label{e032}
N_{\rm o}&=&\frac{1}{2}\left( \langle X_{\rm o}^2 \rangle+\langle Y_{\rm o}^2 \rangle-1\right),
\end{eqnarray}
where $o=a, b,c$ correspond to the excitation numbers of the cavity mode, the Kittel mode, and the HMS mode, respectively. Figures~\ref{fig2}\textcolor{blue}{(e)}-\ref{fig2}\textcolor{blue}{(h)} present the excitation numbers $N_{\rm a}$~[Fig.~\ref{fig2}\textcolor{blue}{(e)}], $N_{\rm b}$~[Fig.~\ref{fig2}\textcolor{blue}{(f)}], $N_{\rm c}$~[Fig.~\ref{fig2}\textcolor{blue}{(g)}], and $N_{\rm b}+N_{\rm c}$~[Fig.~\ref{fig2}\textcolor{blue}{(h)}] as a function of the two detunings of $\tilde{\Delta}_{\rm b}$ and $\Delta_{\rm a}$, where two self-Kerr effects are not considered. The numerical results are carried out in a zero-temperature environment ($T_{\rm e}=0$ K) and a strong coupling regime ($g_{\rm ab},\tilde{G}>\gamma_{\rm a},\gamma_{\rm b},\gamma_{\rm c}$), where the dissipation rate for each mode is chosen from experimentally feasible parameters. All the parameters ensure that the system is always stable. We can find that the frequency ranges of the excited two magnon modes are complementary [see Figs.~\ref{fig2}\textcolor{blue}{(f)} and~\ref{fig2}\textcolor{blue}{(g)}]. $N_{\rm a}\simeq N_{\rm b}+N_{\rm c}$ implies that no matter which magnon mode is excited, it is accompanied by the scattering of microwave photons [see Figs.~\ref{fig2}\textcolor{blue}{(e)} and~\ref{fig2}\textcolor{blue}{(h)}]. The schematic diagram corresponding to the numerical results in Fig.~\ref{fig2} is shown in Figs.~\ref{fig1}\textcolor{blue}{(b)} and~\ref{fig1}\textcolor{blue}{(c)}, where the matching condition of parametric conversion process determines the optimal frequency detunings at which cavity photons can be entangled.


\section{Self-Kerr effect induced entangled state transfer}\label{s5}

In fact, it is unavoidable that the effective self-Kerr effects will also be enhanced when the YIG sphere is pumped by the drive field. This is a problem that the majority of entanglement generation schemes avoid, namely the presence of multiple nonlinearities in the system. Numerous nonlinear combined effects contribute to the system's difficulty of analysis and instability. However, we demonstrate in this article that the presence of self-Kerr effects enhances magnon-magnon entanglement transfer into other subsystems, resulting in a significant increase in tripartite entanglements.

\begin{figure}[b]
	\hskip-0.19cm\includegraphics[width=\linewidth]{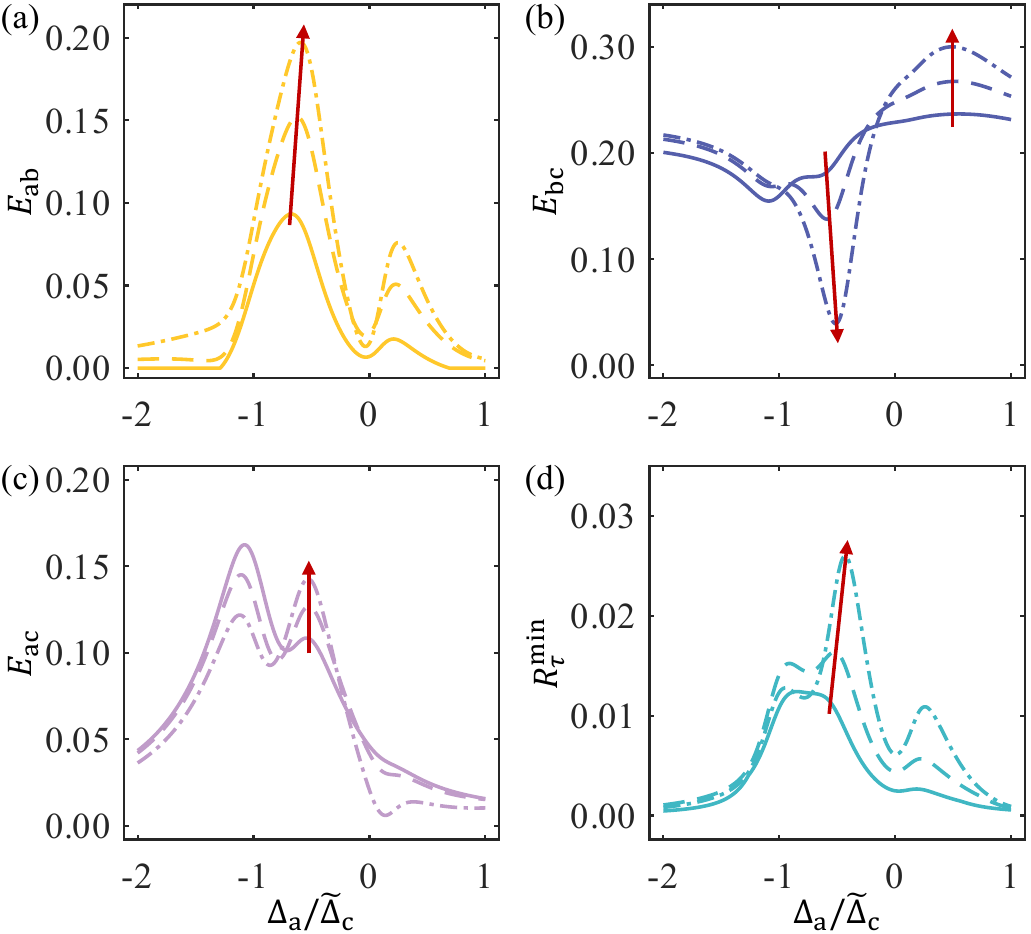}
	\caption{Bipartite entanglements (a) $E_{\rm ab}$, (b) $E_{\rm bc}$, (c) $E_{\rm ac}$ and tripartite entanglement (d) $R_{\tau}^{\rm min}$ versus $\Delta_{\rm a}$ for different self-Kerr coefficients $\tilde{K}_{\rm b}$ and $\tilde{K}_{\rm c}$. We take $\tilde{K}_{\rm b}=\tilde{K}_{\rm c}=0$ (solid); $\tilde{K}_{\rm b}/2\pi=7.5$ MHz, $\tilde{K}_{\rm c}/2\pi=12$ MHz (dashed); $\tilde{K}_{\rm b}/2\pi=15$ MHz, $\tilde{K}_{\rm c}/2\pi=24$ MHz (dot-dashed). The arrows depict the evolution of the entanglement as the two self-Kerr effects are enhanced. The other parameters are listed in Tab.~\ref{tab1}.}
	\label{fig3}
\end{figure}
Figure~\ref{fig3} describes bipartite entanglements $E_{\rm ab}$~[Fig.~\ref{fig3}\textcolor{blue}{(a)}], $E_{\rm bc}$~[Fig.~\ref{fig3}\textcolor{blue}{(b)}], $E_{\rm ac}$~[Fig.~\ref{fig3}\textcolor{blue}{(c)}] and tripartite entanglement $R_{\tau}^{\rm min}$~[Fig.~\ref{fig3}\textcolor{blue}{(d)}] as a function of the detuning $\Delta_{\rm a}$ for different self-Kerr coefficients, where a smaller cavity mode-Kittel mode coupling rate $g_{\rm ab}/2\pi=30$ MHz is chosen for better illustration. Figure~\ref{fig3}\textcolor{blue}{(b)} shows that when $\Delta_{\rm a}\simeq-\tilde{\Delta}_{\rm c}$, $E_{\rm bc}$ decreases gradually as two self-Kerr coefficients increase. Instead, $E_{\rm ab}$ and $E_{\rm ac}$ gradually increase in the process [see Figs.~\ref{fig3}\textcolor{blue}{(a)} and~\ref{fig3}\textcolor{blue}{(c)}], which implies that more entanglement in Kittel mode-HMS mode subsystem is transferred to the cavity mode-Kittel mode and cavity mode-HMS mode subsystems. The tripartite entanglement in terms of the minimum residual contangle becomes stronger when the entanglement is more evenly distributed in each subsystem, as illustrated in Fig.~\ref{fig3}\textcolor{blue}{(d)}.

\begin{figure}[t]
	\hskip-0.19cm\includegraphics[width=\linewidth]{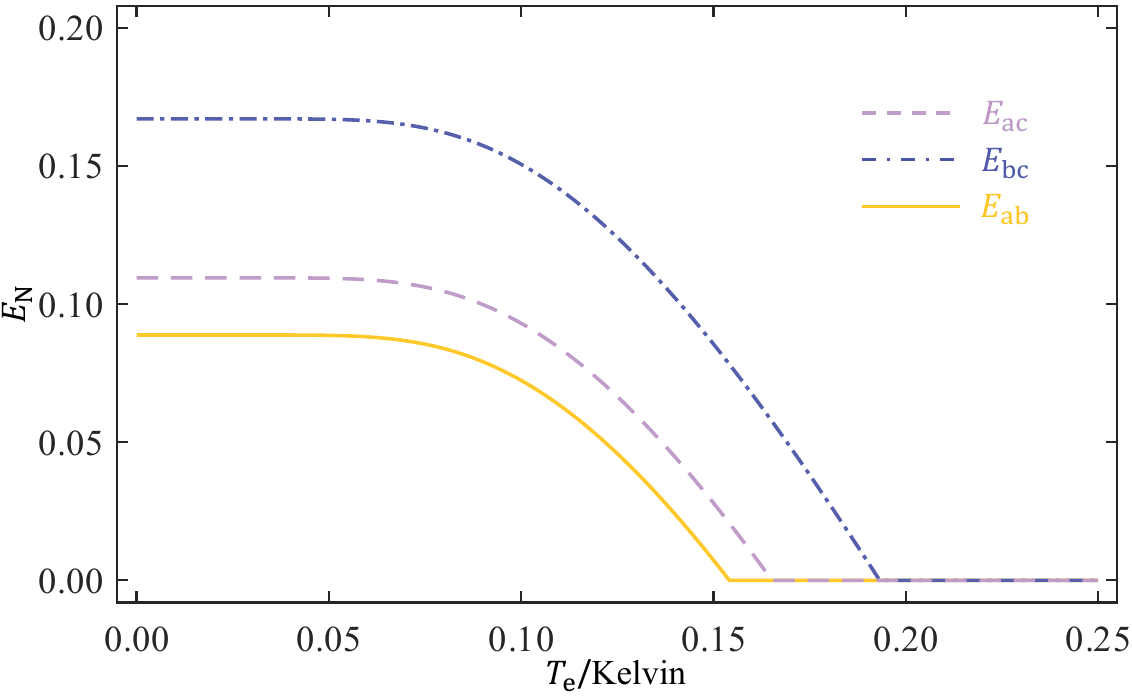}
	\caption{Bipartite entanglements $E_{\rm ab}$, $E_{\rm bc}$, and $E_{\rm ac}$ versus environmental temperature $T_{\rm e}$. See Tab.~\ref{tab1} for the the parameters.}
	\label{fig4}
\end{figure}

When the two self-Kerr effects are considered, the linearized Hamiltonian of the system can be rewritten as
\begin{eqnarray}\label{e033}
H_{\rm L}/\hbar&=&\Delta_{\rm a}a^\dag a+\tilde{\Delta}_{\rm b}b^\dag b+\tilde{\Delta}_{\rm c}c^\dag c+g_{\rm ab}(a^\dag b+ab^\dag)\nonumber\\
&&+\tilde{K}_{\rm b}(b^{\dag}b^{\dag}+bb)/2+\tilde{K}_{\rm c}(c^{\dag}c^{\dag}+cc)/2\nonumber\\
&&+\tilde{G}(b^\dag+b)(c^\dag+c).
\end{eqnarray}
To show the mechanism, we diagonalize these two terms by introducing squeezing operators
\begin{eqnarray}\label{e034}
S(\theta_{\rm b})&=&\exp\left[ \theta_{\rm b}(bb-b^{\dag}b^{\dag})\right] ,\nonumber\\
S(\theta_{\rm c})&=&\exp\left[ \theta_{\rm c}(cc-c^{\dag}c^{\dag})\right] .
\end{eqnarray}
The two Bogoliubov modes can be written as
\begin{eqnarray}\label{e035}
\beta_{\rm b}&=&S^{\dag}(\theta_{\rm b})bS(\theta_{\rm b})=\cosh\theta_{\rm b}b-\sinh\theta_{\rm b}b^{\dag},\nonumber\\
\beta_{\rm c}&=&S^{\dag}(\theta_{\rm c})cS(\theta_{\rm c})=\cosh\theta_{\rm c}c-\sinh\theta_{\rm c}c^{\dag},
\end{eqnarray}
where
\begin{eqnarray}\label{e036}
\theta_{\rm b(c)}&=&\frac{1}{4}\ln\mathcal{C}_{\rm b(c)}~{\rm with}~\mathcal{C}_{\rm b(c)}=\frac{\tilde{\Delta}_{\rm b(c)}-\tilde{K}_{\rm b(c)}}{\tilde{\Delta}_{\rm b(c)}+\tilde{K}_{\rm b(c)}}.
\end{eqnarray}
For simplicity, we set $\tilde{\Delta}_{\rm b}\simeq\tilde{\Delta}_{\rm c}\simeq\tilde{\Delta}$ and $\tilde{K}_{\rm b}\simeq\tilde{K}_{\rm c}\simeq\tilde{K}$, so $\mathcal{C}_{\rm b}\simeq\mathcal{C}_{\rm c}\simeq\mathcal{C}$ and $\theta_{\rm b}\simeq\theta_{\rm c}\simeq\theta$. In fact, $\arrowvert\tilde{\Delta}_{\rm b}\arrowvert<\arrowvert\tilde{\Delta}_{\rm c}\arrowvert$ and $\tilde{K}_{\rm b}<\tilde{K}_{\rm c}$, but the above setting does not lose the generality of the analysis. Therefore, the Bogoliubov Hamiltonian can be written as
\begin{eqnarray}\label{e037}
H_{\rm B}/\hbar&=&\Delta_{\rm a}a^\dag a+\Delta_{\beta}\beta_{\rm b}^{\dag}\beta_{\rm b}+\Delta_{\beta}\beta_{\rm c}^{\dag}\beta_{\rm c}+\mathcal{G}(\beta_{\rm b}+\beta_{\rm b}^{\dag})(\beta_{\rm c}+\beta_{\rm c}^{\dag})\nonumber\\
	&&+g_{\rm ab}^{\rm cos}(\beta_{\rm b}^{\dag}a+\beta_{\rm b}a^{\dag})+g_{\rm ab}^{\rm sin}(\beta_{\rm b}^{\dag}a^{\dag}+\beta_{\rm b}a),
\end{eqnarray}
where $g_{\rm ab}^{\rm cos}=g_{\rm ab}\cosh\theta$, $g_{\rm ab}^{\rm sin}=g_{\rm ab}\sinh\theta$, $\Delta_{\beta}=(\tilde{\Delta}^2-\tilde{K}^2)^{\frac{1}{2}}$, and $\mathcal{G}=\tilde{G}\sqrt{\mathcal{C}}$. Equation~(\ref{e037}) shows that $\Delta_{\rm a}\simeq-\Delta_{\beta}$ is optimal for the cavity mode-Kittel mode entanglement, due to the squeezing term $g_{\rm ab}^{\rm sin}(\beta_{\rm b}^{\dag}a^{\dag}+\beta_{\rm b}a)$, which also results in a frequency shift for optimal detuning. In fact, when self-Kerr nonlinearity is introduced, the entanglement $E_{\rm bc}$ becomes stronger because $\mathcal{C}>1$ [see Fig.~\ref{fig3}\textcolor{blue}{(b)} for $\Delta_{\rm a}\not\simeq-\tilde{\Delta}_{\rm c}$]. At the same time, the cavity mode-Kittel mode state-swap interaction is also enhanced because $g_{\rm ab}^{\rm cos}>g_{\rm ab}$ for $\theta\not=0$. Thus, the enhancement of $E_{\rm ab}$ and $E_{\rm ac}$ is caused by two reasons: (i) more entanglement is transferred into the subsystem containing the cavity mode [see Figs.~\ref{fig3}\textcolor{blue}{(a)}-\ref{fig3}\textcolor{blue}{(c)}]; (ii) the emergence of a new two-mode squeezing term $g_{\rm ab}^{\rm sin}(\beta_{\rm b}^{\dag}a^{\dag}+\beta_{\rm b}a)$. The results show that the self-Kerr effect facilitates the transfer of entanglement, which can make the minimum residual contangle $R_{\tau}^{\rm min}$ larger than the previous value [see Fig.~\ref{fig3}\textcolor{blue}{(d)} for $\Delta_{\rm a}\simeq-\tilde{\Delta}_{\rm c}$]. Finally, the generated subsystem entanglements are robust against environmental temperature and the maximum survival temperature (MST) is about 0.15$\sim$0.2 Kelvin, as shown in Fig.~\ref{fig4}.

\section{Entanglement detection and application}\label{s6}

The generated magnon-magnon entanglement can be detected by measuring the quadratures of the two magnon modes $X_{\rm b/c}$ and $Y_{\rm b/c}$, and then calculating the covariance matrix. The entanglement parameter region shown Fig.~\ref{fig2} shows that entanglement can be obtained even when there is certain frequency detuning between the two magnon modes. Two weak microwave probe fields resonantly coupled to the two detuned magnon modes can read out the four quadratures by using the cavity-magnon beam-splitter interaction. Here, we focus on the detection of magnon-magnon entanglement $E_{\rm bc}$. To achieve the entanglement detection, the dissipation rate of two magnon modes should be much lower than that of the cavity mode (e.g., we take $\gamma_{\rm a}/2\pi=18.6$ MHz and $\gamma_{\rm b/c}/2\pi=6.7$ MHz in Fig.~\ref{fig3}). As a result, when the pump tone is turned off, cavity microwave photons dissipate quickly. We send the two probe fields after the cavity photons dissipate completely. Then, the probe output fields contain only the information regarding the entanglement of the two magnon modes.


Quantum entanglement is a phenomenon wherein systems cannot be described independently of one other despite being separated by an arbitrarily large distance. It is also the key resource behind many emerging quantum technologies, such as quantum computing~\cite{RR-01,EK-01,Ladd-10,Lidar-18} and metrology~\cite{Lloyd-06,Lloyd-11}. The entanglement of two different degrees of freedom inside one ferrimagnetic crystal provides a concept for CV information processing at the mesoscopic scale. Our research sheds light on the entanglement scheme between additional HMS modes and Kittel mode induced by their nonlinear couplings.

\section{Conclusions and perspectives}\label{s7}
In summary, we have presented a scheme to generate steady-state entanglement in a cavity magnonic system where a microwave cavity mode is coupled to a Kittel mode in a YIG sphere, and the Kittel mode is simultaneously coupled to a HMS mode via mode overlap, which originates from the partial local spins shared by the Kittel mode and other spin wave modes~\cite{AG-19,DDS-09}. In such a system, we study the properties of entangled magnon modes, and find that the cross-Kerr effect is able to induce steady-state entanglement between two magnon modes with experimentally accessible parameters. Additionally, when the spontaneous parametric process occurs, the cavity photons also become entangled with the magnons. We also demonstrate the effect of the self-Kerr nonlinearities on the bipartite and tripartite entanglements, where the mutual coupling between different modes becomes stronger and a new two-mode squeezed state is generated. When the two types of nonlinearities coexist, the entanglement is more uniformly distributed across the subsystems, and the tripartite entanglement is also enhanced.

Our work will open up new avenues for studying entanglement when multiple nonlinearities exist, as well as for realizing an entangled state within a single YIG sphere, which will enable spatially localized conservation of entanglement in ferrimagnetic spin ensembles. Moreover, the method for generating steady-state entanglement via the cross-Kerr effect could be extended to other hybrid systems. In the quantum magnonic systems~\cite{Nakamura-19}, a cross-Kerr nonlinear interaction between the magnetostatic mode and the qubit is also facilitated by their mutual couplings to microwave cavity modes~\cite{YH-11,ETH-15}, which provides the additional nonlinearity required to investigate quantum effects in magnonics.

\begin{acknowledgments}
This work is supported by the National Natural Science Foundation of China (No.~$11934010$, No.~${\rm U}1801661$, and No.~$12174329$), Zhejiang Province Program for Science and Technology (No.~$2020{\rm C}01019$), and the Fundamental Research Funds for the Central Universities (No.~$2021{\rm FZZX}001$-$02$).
\end{acknowledgments}

\appendix
\section{The quantification of entanglements}\label{appA}

Here we briefly give the quantification of the bipartite and tripartite entanglements. To study the bipartite CV entanglements, we introduce the logarithmic negativity, which is defined as~\cite{GV-02,MBP-05}
\begin{eqnarray}\label{e026}
	E_{\rm N}=\max\left[ 0,-\ln\left( 2\nu^{-}\right) \right],
\end{eqnarray}
where $\nu^{-}=\min\arrowvert{\rm eig}\oplus_{\rm s=1}^2 (-\sigma_{\rm y}) P_{1\arrowvert2}V_{4}P_{1\arrowvert2}\arrowvert$ with $\sigma_{\rm y}$ as the $\rm y$-Pauli matrix. $V_{4}$ is the $4\times4$ CM of the two subsystems that only includes the rows and columns of the interested modes in $V$, and the matrix $P_{1\arrowvert2}=\sigma_{\rm z}\oplus\mathbb{1}$ (with the identity matrix $\mathbb{1}$) realizes partial transposition at the CM level. In the main text, we use $E_{\rm ab}$, $E_{\rm ac}$, and $E_{\rm bc}$ to denote the cavity mode-Kittel mode, the cavity mode-HMS mode, and the Kittel mode-HMS mode entanglements, respectively. 

To investigate the tripartite CV entanglement, we introduce residual contangle defined as~\cite{GA-06,GA-07}
\begin{eqnarray}\label{e027}
	R_{\tau}^{\rm i\arrowvert jk}=C_{\rm i\arrowvert jk}-C_{\rm i\arrowvert j}-C_{\rm i\arrowvert k}
\end{eqnarray}
with $i,j,k=a,b,c$. $C_{\rm m\arrowvert n}=E_{\rm m\arrowvert n}^2$, as the squared logarithmic negativity with entanglement monotonicity, is the contangle of $m$ and $n$ subsystems, where $n$ may involve one or two modes. The single-mode versus dual-mode logarithmic negativity is defined as 
\begin{eqnarray}\label{e028}
	E_{\rm i\arrowvert jk}=\max\left[0,-\ln\left( 2\nu^{-}_{\rm i\arrowvert jk}\right) \right],
\end{eqnarray}
where $\nu^{-}_{\rm i\arrowvert jk}=\min\arrowvert{\rm eig}i\oplus_{\rm s=1}^{3}(i\sigma_{\rm y})\tilde{V}\arrowvert$ is the minimum symplectic eigenvalue of the $6\times6$ CM $\tilde{V}=P_{\rm i\arrowvert jk}VP_{\rm i\arrowvert jk}$. The matrices $P_{1\arrowvert23}=\sigma_{\rm z}\oplus\mathbb{1}\oplus\mathbb{1}$, $P_{2\arrowvert13}=\mathbb{1}\oplus\sigma_{\rm z}\oplus\mathbb{1}$, and $P_{3\arrowvert12}=\mathbb{1}\oplus\mathbb{1}\oplus\sigma_{\rm z}$ are used for partial transposition at the level of the full CM. $R^{\rm i\arrowvert jk}_{\tau}\ge0$ implies that the residue contangle $R_{\tau}$ satisfies the quantum entanglement monogamy. The minimum residual contangle is defined as~\cite{GA-06,GA-07}
\begin{eqnarray}\label{e029}
	R^{\rm min}_{\tau}=\min[R^{\rm a\arrowvert bc}_{\tau},R^{\rm b\arrowvert ac}_{\tau},R^{\rm c\arrowvert ab}_{\tau}],
\end{eqnarray}
which characterizes a {\it bona fide} three-party property of the CV three-mode Gaussian states.

\end{document}